\documentclass[fleqn,10pt]{wlscirep}
\usepackage{siunitx}
\usepackage{threeparttable}
\title{Integrated and efficient diffusion-relaxometry using ZEBRA}

\author[1,*]{Jana Hutter}
\author[2]{Paddy J Slator}
\author[1]{Daan C Christiaens}
\author[1]{Rui Pedro AG Teixeira}
\author[1]{Thomas Roberts}
\author[1]{Laurence Jackson}
\author[1]{Anthony N Price}
\author[1]{Shaihan Malik}
\author[1]{Joseph V Hajnal}

\affil[1]{School of Biomedical Engineering \& Imaging Sciences, King's College London, London, UK}
\affil[2]{Centre for Medical Image Computing and Department of Computer Science, University College London, London, UK}

\affil[*]{Jana Hutter, Perinatal Imaging and Health, 1st Floor South Wing, St. Thomas' Hospital, Westminster Bridge Road, SE17EH, London, UK, jana.hutter@kcl.ac.uk}


\begin{abstract}
The emergence of multiparametric diffusion models combining diffusion and relaxometry measurements provide powerful new ways to explore tissue microstructure with the potential to provide new insights into tissue structure and function. However, their ability to provide rich analyses and the potential for clinical translation critically depends on the availability of efficient, integrated, multi-dimensional acquisitions. We propose a fully integrated sequence simultaneously sampling the acquisition parameter spaces required for T1 and T2* relaxometry and diffusion MRI. Slice-level interleaved diffusion encoding, multiple spin/gradient echoes and slice-shuffling are combined for higher efficiency, sampling flexibility and enhanced internal consistency. In-vivo data was successfully acquired on healthy adult brains. Obtained parametric maps as well as clustering results demonstrate the potential of the technique regarding its ability to provide eloquent data with an acceleration of roughly 20 compared to conventionally used approaches. The proposed integrated acquisition, called ZEBRA, offers significant acceleration and flexibility compared to existing diffusion-relaxometry studies and thus facilitates wider use of these techniques both for research-driven and clinical applications.\\
\end{abstract}
\begin{document}

\flushbottom
\maketitle
\thispagestyle{empty}

\section*{Introduction}
\label{sec:Intro}

Magnetic Resonance Imaging (MRI) is characterised by its flexible contrast with strong dependencies on multiple diverse tissue parameters, thus enabling sensitizing the measurement to a wide range of properties in healthy tissue and pathology. One key limitation, however, is the intrinsic macroscopic resolution of MRI, while attempting to probe microscopic tissue features. Every voxel is composed of a variety of microstructural compartments with different properties and interactions. There is therefore a growing interest in quantitative MRI in which multiple measurements are combined with a suitable model to facilitate estimating microscopic parameters of interest~\cite{Novikov2018b}. Two main families of techniques in quantitative MRI are diffusion Magnetic Resonance Imaging (dMRI)~\cite{LeBihan1986} and MR relaxometry.\\

\noindent The classical approach is to seek quantifying one parameter or family of tissue parameters from one modality at a time. This has the benefits of limiting the complexity of the modelling required and reducing the amount of data that must be acquired, which is often an over-riding concern for in-vivo studies, particularly if clinical applicability is sought. However, joint analyses attempting to quantify multiple tissue parameters from multiple modalities at the same time have great potential, particularly if diverse physical processes are involved.\\

\noindent Diffusion MRI is a powerful modality for probing the microstructural architecture of biological tissues in-vivo~\cite{LeBihan1986,Callaghan1988}. The applied diffusion encoding gradients render this technique sensitive to the random motion of water molecules within the sample---thus informing on the underlying microstructure~\cite{Beaulieu2002}. Varying the diffusion encoding strength ($b$-value) and direction ($b$-vector) extracts information at different length scales and orientations, hence enabling to estimate microstructural compartment properties and neural fibre orientations~\cite{Jespersen2007,Tournier2007,Reisert2017}. Recent work has highlighted the limitations of microstructure imaging from dMRI alone. For example, disentangling intra- and extra-axonal compartment properties in brain white matter is degenerate and ill-posed~\cite{Novikov2018a}. An emerging trend is therefore to combine dMRI with other contrasts to disentangle multi-compartmental effects~\cite{DeSantis2014,Veraart2017}.\\

\noindent Recent work~\cite{Novikov2016} highlighted the limitations of microstructural imaging from dMRI alone. For example, disentangling intra- and extra-axonal compartment properties in brain white matter is degenerate and ill-posed. An emerging trend is the combination of dMRI with other contrasts to disentangle multi-compartmental effects.~\cite{DeSantis2014,Veraart2017}.\\

\noindent Relaxometry exploits the inherent sensitivity of MRI to the biochemical environment of tissue, captured by transverse (T2 and T2$^*$) and longitudinal (T1) relaxation times. T1 processes describe the recovery of longitudinal magnetization while T2 processes describe the loss of transverse magnetization due to dephasing effects originating from interactions at the molecular level. Additional dephasing---caused mainly by field inhomogeneities, differences in susceptibility, and chemical shift effects---are described by T2', contributing to the shorter characteristic T2$^*$ time. Thus, changes in T1, T2 and T2$^*$directly reflect the biochemical environment in tissue. Relaxometry has been fundamental to enabling quantitative measurements of tissue properties in health and disease and has been proven in a wide variety of tissues.\\ 

\noindent Biophysically-linked tissue parameters which have been extracted from in-vivo relaxometry or dMRI data include but are not limited to: axon density \cite{DeSantis2014,Lutti2014}, myelin fraction \cite{Koenig1990}, deoxyhemoglobin concentration, iron concentration \cite{Deoni2010,Liu2009} and biliary fibrosis \cite{Gao2016}. While novel relaxometry techniques such as MR Fingerprinting \cite{Ma2013} or steady-state gradient echo based methods \cite{Teixeira2018,Deoni2003,Marques2010,Weiskopf2013} are available, the accepted gold-standard techniques are still inversion-recovery (IR) sequences for T1 mapping, and repetitions with multiple echo times for T2/T2$^*$ mapping.  IR sequences sample the T1 domain by varying the inversion time (TI) between the inversion pulse and the acquisition, while the acquisition of multiple echoes samples the time between excitation and acquisition (TE) \cite{Deoni2010}. Repeated gradient-echo experiments allow T2$^*$ quantification, whereas repeated spin-echo experiments estimate T2 - since  the refocusing pulse eliminates the previously discussed  sources of  additional transverse magnetisation decay (i.e. T2').\\

\noindent Joint relaxometry-diffusion experiments have been conducted successfully in the context of nuclear magnetic resonance (NMR) spectroscopy, improving the ability the distinguish different compartments \cite{Bernin2013,DeAlmeidaMartins2018}. The recent study by Almeida et al. \cite{DeAlmeidaMartins2018} samples the three-dimensional space T1-T2-diffusion, and adds the shape of the diffusion encoding tensor to bval/\textbf{bvec} as an additional dimension within the diffusion parameters. Recent studies saw the extension of these techniques to imaging, Kim et al. \cite{Kim2017} and Benjamini et al. \cite{Benjamini2017} illustrated increased separation of compartments in injured spinal cord tissue by sampling the T2-diffusion and T1-T2-diffusion spaces respectively.\\

\begin{figure}[t]
  \centering
  \scriptsize
  \includegraphics[width=0.99\textwidth]{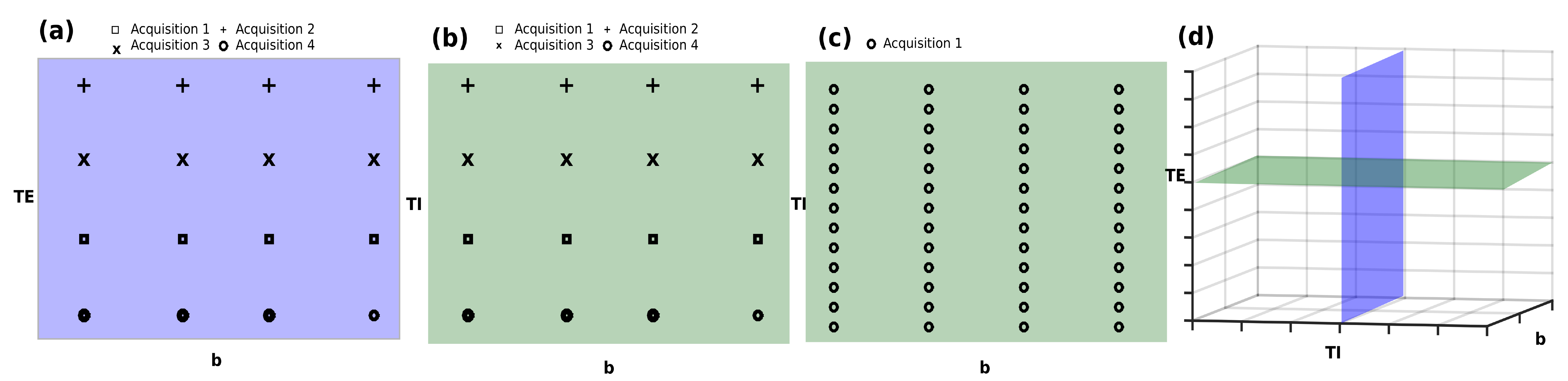}
  \caption{Sampling schemes of the multidimensional acquisition parameter space. (a) Scheme of conventional T2-diffusion acquisitions such as e.g. Veraart et al., sampling the TE-b domain with individual diffusion acquisitions for each considered echo time. (b) Scheme of conventional T1-diffusion acquisitions with separate diffusion acquisitions per TI such as e.g. DeSantis at al. (c) Scheme as employed by T1-diffusion studies using slice-shuffling approaches to sample the TI-diffusion domain in one scan. (d) Depiction of the two considered combinations T1-diffusion and T2-diffusion as planes within the larger three-dimensional TI-TE-b space.}
  \label{fig:sampling_con}
\end{figure}

\noindent In-vivo application of joint relaxometry-diffusion studies has been hampered by the prohibitively long acquisition times. However, recent in-vivo studies by Cherubini et al. \cite{Cherubini2009}, Tax et al. \cite{Tax2017}, and Veraart et al. \cite{Veraart2017} demonstrated joint T2*/T2-diffusion experiments and showed great promise. Additionally Arazany and Assaf \cite{Barazany2012}, and DeSantis et al. \cite{DeSantis2016} have published joint T1-diffusion studies. However, these studies share three significant limitations:
\begin{enumerate}
\item no in-vivo study to date has sampled both longitudinal and transverse relaxation times (Fig. \ref{fig:sampling_con}d)
\item every acquisition parameter combination in the chosen plane (Fig. \ref{fig:sampling_con}a-b) is sampled in a separate acquisition (acquisition is defined in the following by their separate preparation time: for T1 the required TI, for T2$^*$ the required time until the desired TE)
\item all the mentioned studies reported scan times in excess of 60  minutes
\end{enumerate}

\noindent Only challenge (2) has been addressed so far in the context of joint T1-diffusion experiments by the introduction of slice-shuffling (SS) approaches by Ordidge et al. \cite{Ordidge1990} and Wu et al. \cite{Wu2017}. In this approach the slice acquisition order is changed from volume-to-volume to achieve sampling of every spatial location with diffusion weighting at every inversion time TI (see Fig. \ref{fig:sampling_con}c). The nested structure of this approach makes use of all required preparation times for the acquisition of different slices. However, while limiting the preparation idle time, current SS techniques are not optimally efficient: the intrinsic link between the number of inversion times to the number of slices effectively oversamples the TI range for many geometry choices.\\
\noindent Attempts to sample transverse relaxation time are currently done by acquiring multiple repeats of the dMRI or SS acquisition with varying TE.\\

\noindent In this study, we introduce an integrated efficient acquisition technique which can address the three stated challenges and allow efficient sampling of the three-dimensional acquisition parameter space (here T1-T2$^*$-diffusion). This is achieved by nesting the sampling of all required acquisition parameter combinations into one acquisition such that all required preparation times are used. We call it ZEBRA in recognition of the intrinsically striped appearance of slice stacks with echo time, b value and TI all varying . The proposed method is enabled by two novel methodological elements:
\begin{itemize}
\item (I) A multi-echo approach, combining a spin echo with subsequent gradient echoes allows simultaneous exploration of the T2*-diffusion space.
\item (II) Efficient sampling of the inversion recovery curve by interleaving the diffusion encodings in super-blocks, which allows more diverse diffusion weightings to be sampled in the time saved by avoiding over-sampling the TI dimension..
\end{itemize}

\noindent We demonstrate the resulting integrated T1-T2$^*$-Diffusion acquisition in vivo on adult brain data. The obtained quantitative maps show the potential of this technique and test-retest variability experiments demonstrate its robustness. Finally, multi-dimensional scatter plots of the obtained tissue parameters and direction specific T2$^*$ maps give a first hint at the substantial information.

\section*{Experiments}
\noindent The novel ZEBRA sequence as described in the methods section was implemented on a clinical Philips Achieva-Tx 3T scanner (Release 3.2). The presented experiments were obtained on a clinical 3T Philips Achieva scanner using the 32ch adult head coil. All methods were carried out in accordance with relevant guidelines and regulations, healthy volunteer scanning was approved by the Riverside Research Ethics Committee (REC 01/11/12) and informed written consent was obtained prior to imaging. The data was obtained using a 32ch adult head coil. An adiabatic non-slice selective inversion pre-pulse \cite{Tannus1997} was employed for the global inversion.

\begin{table}[bt]
\small
\caption{Performed experiment parameters. IR: Inversion recovery with different TIs in separate scans, MEGE: Multiecho-Gradient echo sequence, IR-dMRI: Inversion-recovery diffusion sequence with Slice-Shuffling, Res: Resolution}
\label{tab:experiments}
\begin{tabular}{l|lll|ll}
\hline
\textbf{E1: Validation} & \textbf{TI/TR} & \textbf{TE} &\textbf{d (\textbf{bval}/bvec)} &  \textbf{Parameter} \\
dMRI&TR=10000&57&0, 333, 667, 1000    &  \\
IR&500,1000,..,5000ms&57&0&    \\
MEGE &TR=10000&57,81,171,228,285&0 &  \\
ZEBRA &50-6500&57,81,171,228,285&0,333,667,1000& \\
\hline
\textbf{E2: Interleave} & \textbf{TI} & \textbf{TE} &\textbf{d (\textbf{bval}/bvec)} &  \textbf{Parameter} \\
ZEBRA&50-6500& 57& 0  &   Res. 2.6mm$^3$\\
\hline
\textbf{E3: Test-Retest} & \textbf{TI (ms)} & \textbf{}{TE (ms)} &\textbf{d (\textbf{bval}/bvec) (mm$^{-2}$s)} & \textbf{Parameter} \\
ZEBRA&50-6500& 57, 81, 171, 228, 285& 0(1),333(1),667(1),1000(1)   &  Res. 2.6mm$^3$\\
\hline
\textbf{E4: ZEBRA} & \textbf{TI (ms)} & \textbf{TE (ms)} &\textbf{d (\textbf{bval}/bvec) (mm$^{-2}$s)} &  \textbf{Parameter} \\ &50-6500& 57, 81, 171, 228, 285& 0(1),333(1),667(1),1000(1)   &  Res. 2.6mm$^3$\\
\hline
\textbf{E5: HARDI-ZEBRA} & \textbf{TI (ms)} & \textbf{TE (ms)} &\textbf{d (\textbf{bval}/bvec) (mm$^{-2}$s)} &  \textbf{Parameter} \\
&8000& 65, 130, 205, 270 & 0(4),500(6),1000(8),2600 (24)&  Res. 2.6mm$^3$\\
\hline  % Please only put a hline at the end of the table
\end{tabular}
\end{table}

\subsection*{Validation experiment}
Phantom validation was performed by imaging an in-house built spherical phantom filled with a 2\% agarose solution and primarily focused on the goal to comparing the tissue parameter estimates obtained via the ZEBRA approach to the conventional approaches employed for estimating $T2^*$, $T1$ and $ADC$.\\

\noindent Reference T2$^*$ values were found via a multi-echo gradient echo sequence (MEGE) sequence. A gold-standard IR sequence was employed for T1 measurements and a conventional SE dMRI sequence was acquired in order to estimate ADC. All acquisitions used single-shot EPI, with a fixed FOV of 260x260mm$^{2}$, resolution 3mm$^{3}$. The MEGE, dMRI and ZEBRA acquisitions had $N_s=28$ slices, except for IR, where only one slice was acquired to keep the imaging time acceptable. See Table \ref{tab:experiments} for all acquisition parameter details. The ZEBRA data was processed individually to match the equivalent processing of the individually acquired data sets: the longest inversion time at b=0 to estimate the T2$^*$ (matching the MEGE data), the longest inversion time at the shortest echo time with all four b-values to estimate the ADC value (matching the dMRI data) and finally all inversion times for b=0 at the shortest echo time to estimate T1 (matching the individual IR experiments).\\

\subsection*{Cramer-Rao Lower Bound Simulations}
To further evaluate the super-block strategy regarding the relation between super-block length and robustness of tissue parameter estimation, simulation experiments were performed. For this, the Cramer-Rao Lower Bound (CRLB) \cite{Teixeira2018} was used on a slice level to predict the minimum obtainable variance of the model parameters given the performed $N_v$ independent measurements and a fixed noise level.\\

\noindent By considering the signal model $g(\mathbf{x},\mathbf{\theta})$, where  $\mathbf{x}=[\alpha_1,\alpha_2,...,\alpha_{N_v}]$ are the user controlled parameter vector $\alpha_i=$ [bval$_{i}$, TI$_{i}$, TE$_{i}$] and $\Theta=[\theta_1,..,\theta_M]$ are the model parameters - here $\theta=[PD, T1, T2^*, ADC, IE]$. For independent measurements following a Gaussian noise distribution with standard deviation $\sigma$, the CRLB is calculated as $(\mathbf{J'}\Sigma \mathbf{J} )^{-1}$, where $\mathbf{J}$ is the jacobian of the signal model obtained as 
\begin{equation}
\begin{bmatrix}
    \frac{\delta g_1}{\delta \theta_j}       & \dots & \frac{\delta g_1}{\delta \theta_M} \\
    \hdotsfor{3} \\
    \frac{\delta g_{N_v}}{\delta \theta_j}       & \dots & \frac{\delta g_{N_v}}{\delta \theta_M} \\
\end{bmatrix}.
\end{equation}
Finally, each parameter is weighted by the inverse square of the assumed tissue parameter value $\Theta_0$ and summed to obtained the precision $p$.\\

\noindent The used model here was the signal decay equation in Eq. \ref{eq:signaldecay}, $\Theta_0$ was chosen as \begin{equation}[PD=1000,T1=1500ms,ADC=0.001mm^{2}s^{-1},T2^*=200ms,IE=2].
\end{equation}

\noindent Other acquisition parameters included $N_s=40$, TR=10s and SNR=200. Considered TIs are $[50:TR/N_s:TR-50]$ and b-values between $0mm^{-2}s$ and $3000mm^{-2}s$ sampled in steps of $250mm^{-2}s$ were included. To evaluate specifically the super-block length, the TE was kept constant at TE=60ms and for the final evaluation only [$S_0$, $T_1$ and $IE$] were included. T2$^{*}$ could be neglected due to the fixed TE and ADC could be neglected due to the fixed b-value range. Each simulated set of measurements thus varied in length, b-value and employed TIs as follows: The TIs were chosen as a regularly spaced subset of the considered TIs $[50:TR/N_s:TR-50]$. For each scenario, both the obtained precision $p$ and the number of required volumes $N_v$ (and thus required time) were evaluated. The goals therefore are twofold: Quantitatively establishing the influence of the super block length for a range of b-values and illustrating how the flexible approach for combining different super-block length within one acquisition can help standardise the obtained variance across all considered b-values.

\subsection*{Super-block length experiment}
\noindent To inform the right choice of super-block length, another experiment (\textbf{E2}) was performed. Here, an IR-DRMI sequence using SS, but without the proposed interleaving was used to acquire an highly sampled inversion curve data (28 points along the inversion recovery curve between 57 and 6500ms) on a single healthy adult brain. From this dataset, subsets with lower TI sampling density were extracted by choosing the super-block length to be $N_i=1,..,10$ and the respective T1 maps calculated.\\

\subsection*{Test-retest experiment}
\noindent A healthy adult was scanned twice with ZEBRA (experiment \textbf{E3}) using the acquisition parameters as specified in table \ref{tab:experiments} to compare test-retest variability. The described fitting algorithm was employed on both datasets, a brain mask calculated (MRTRIX3) and any bulk motion corrected on a volume level using FLIRT \cite{Jenkinson2002}. The ADC, T2$^*$ and T1 values of all voxels within the brain mask were analysed. 

\subsection*{Healthy adult experiments}
\noindent To illustrate the versatility of the proposed approach, two different ZEBRA parametrizations were employed on a total of four adult volunteers:\\

\noindent Two adults were scanned with a protocol sampling only four diffusion encodings $N_d=4$, corresponding to one direction on b=0, b=333, b=667 and b=1000 at $N_e=5$ echo times with one super-block with an interleave factor of $N_i=4$, resulting in $N_{TI}=N_s/N_i=7$ TIs.
Further acquisition parameters are given in table \ref{tab:experiments} E4.\\
\noindent Another 2 adults were scanned with a high angular resolution multi-shell protocol, composed of 3 shells, with 24 directions (b2600), 8 directions (b1000), 6 directions (b500) and 4 b0 volumes. These in total $N_d=42$ directions were scanned in 6 super-blocks, each composed of $N_i=7$ encodings. This interleave factor was chosen based on the CRLB simulations for the highest considered b-value and led to $N_{TI}=N_s/N_i=4$ TI points sampled. The number of echos was fixed as $N_e=4$. Further acquisition parameters are given in table \ref{tab:experiments} E5.\\

\section*{Results}

\subsection*{Validation result}
\noindent The results from the phantom experiment acquired with the different sequence variations together with the acquisition times are given in Table \ref{tab:phantomResults}. The obtained tissue parameters with ZEBRA are very close to those obtained using the conventional methods. The times for the conventional inversion recovery sequence are for one individual slice only, times for dMRI, MEGE and ZEBRA are for the whole volume. \\

\noindent For the entire 'diffusion encoding-TI-TE space' presented here (5 TEs, 7 TIs and 4 diffusion encodings), the reduction achieved by ZEBRA compared with the current most efficient state-of-the art technique (slice-shuffling) equals 20. This results from a 5-fold acceleration by packing the 5 TEs in subsequent echos and by the achieved 4-fold acceleration due to interleaving $N_i=4$ diffusion encodings. All considered conventional and ZEBRA sampling schemes are shown in Fig. \ref{fig:interleaveTest}a.

\subsection*{Cramer-Rao Lower Bound Simulation results}
\noindent The results of the CRLB simulations are given in Fig. \ref{fig:interleaveTest}b. The CRLB for the considered bval-$N_i$ grid (where the number of interleaved bval/bvec combinations $N_i$ defines acceleration) is displayed using a color code: Yellow being the highest variance and thus lowest precision and blue the lowest variance and thus highest precision. The isolines illustrate the required interleave factor across considered b-values to obtain similar variance. For a variance level of 0.1 for example, the highest considered bval of $b=3000mm^{-2}s$ could thus only be accelerated by factor 3, while a $b=0mm^{-2}s$ could be acquired with $N_i=7$ to achieve similar precision in the estimation of [PD,T1,IE] (both marked with a cross in Fig. \ref{fig:interleaveTest}b). Each interleave factor $N_i$ results in a different acquisition time, where $N_i=1$ corresponds to no acceleration and $N_i=10$ to a 10-fold acceleration.

\subsection*{Superblock length experiments}
\begin{figure}[t]
  \centering
  \scriptsize
  \includegraphics[width=0.7\textwidth]{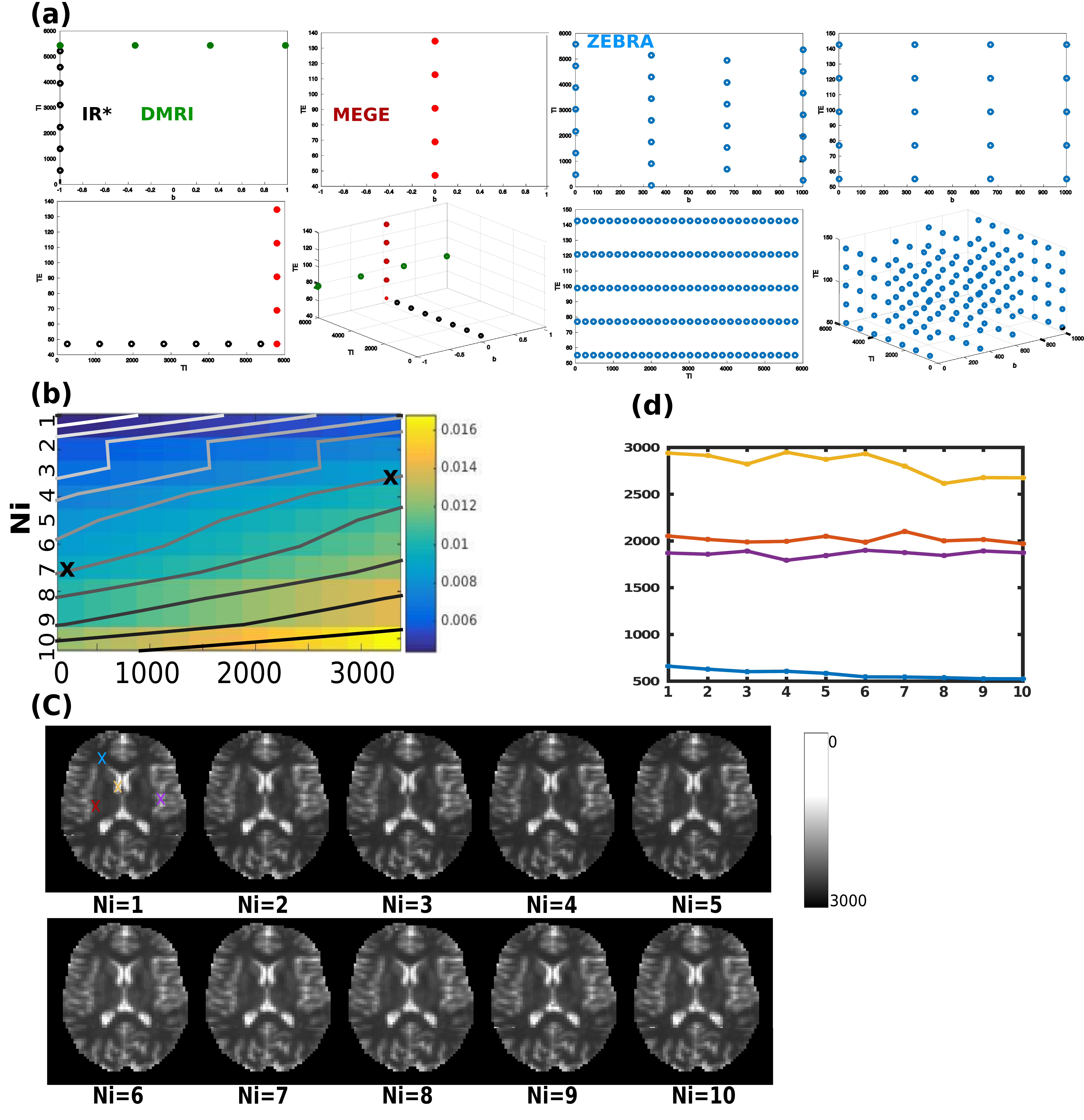}
  \caption{(a) Achieved sampling of the acquisition parameter space with conventional methods and ZEBRA. (b) CRLB results for b-values between 0 and 3000, sampled at 250mm$^{-2}$s and $N_i=1,..,10$ with overlayed contour lines. (c) Adult experiment sampled with 27 TIs - sub-sampled at $N_i=1,..,10$ resulting in 27 to 3 points. The crosses indicate the regions for (d) - showing the obtained T1 values for all interleave factors. (e) A mid-brain slice of the obtained T1 maps from the described adult experiment for all interleave factors in shown - illustrating little variability.}
  \label{fig:interleaveTest}
\end{figure}

\noindent The results from imaging experiment \textbf{(E2)} are depicted in Fig. \ref{fig:interleaveTest}c and d, with\ref{fig:interleaveTest}c showing T1 maps for a selected mid-brain slice. On visual inspection, the results are stable for $N_i=1,..,10$ corresponding to up to 10-fold acceleration. Example T1 values plotted as a function of super-block size, $N_i$, for CSF, grey matter and white matter, sampled at the voxels marked by coloured crosses are plotted in (d). These illustrate stable TI values of $<2$\% difference between $N_i=1$ and $N_i=10$ for TIs in the grey matter voxels and $<5$\% difference for white matter. Only the values in the CSF decline by about 12 \% for $N_i>6$.

\begin{table}[bt]
\small
\caption{Results phantom experiments. IR: Inversion recovery with different TIs in separate scans, MEGE: Multiecho-Gradient echo sequence, d: diffusion encoding combination.}
\label{tab:phantomResults}
\begin{tabular}{llllccc}
\hline 
\textbf{Experiments} & \textbf{Samples} &\textbf{Slices} &\textbf{Acquisition time} & \textbf{T1} &\textbf{ T2*} &\textbf{ADC} \\
\hline
dMRI &4 ($4 d \cdot 1TR \cdot 1TE$)& 28 &0:32& &&0.064 $\pm$ 0.001  \\
IR &7 ($7 TI \cdot 1TE \cdot 1d=b0$)&1&7 $\times$ 0:30 & 2734$\pm$ 9.20&& \\
MEGE &5 ($5 TE \cdot 1TR \cdot 1d =b0$)&28&0:26& &55.12$\pm$3.94&\\
\hline
ZEBRA &140 ($7 TI \cdot 5TE \cdot 4 d$)&28&3:24& 2730$\pm$35.22&54.94$\pm$2.29&0.064$\pm$ 0.001 \\
%\hiderowcolors
%stop alternating row colors from here onwards\\
\hline  % Please only put a hline at the end of the table
\end{tabular}
\end{table}

\subsection*{Test-retest experiments}
\noindent The Bland-Altman plots from the test-retest adult experiment in Supporting Fig. S2 illustrate the ADC, T2$^*$ and T1 for a random subset of all voxels in the brain mask. They show a high degree of reproducibility with $r^2=0.95, 0.96$ and $0.89$ respectively. The relatively high spread of the T1 results is a consequence of subvoxel motion between the scans leading to increased errors, mainly in the CSF.\\

\subsection*{Adult brain results}
\begin{figure}[t]
  \centering
  \scriptsize
  \includegraphics[width=0.9\textwidth]{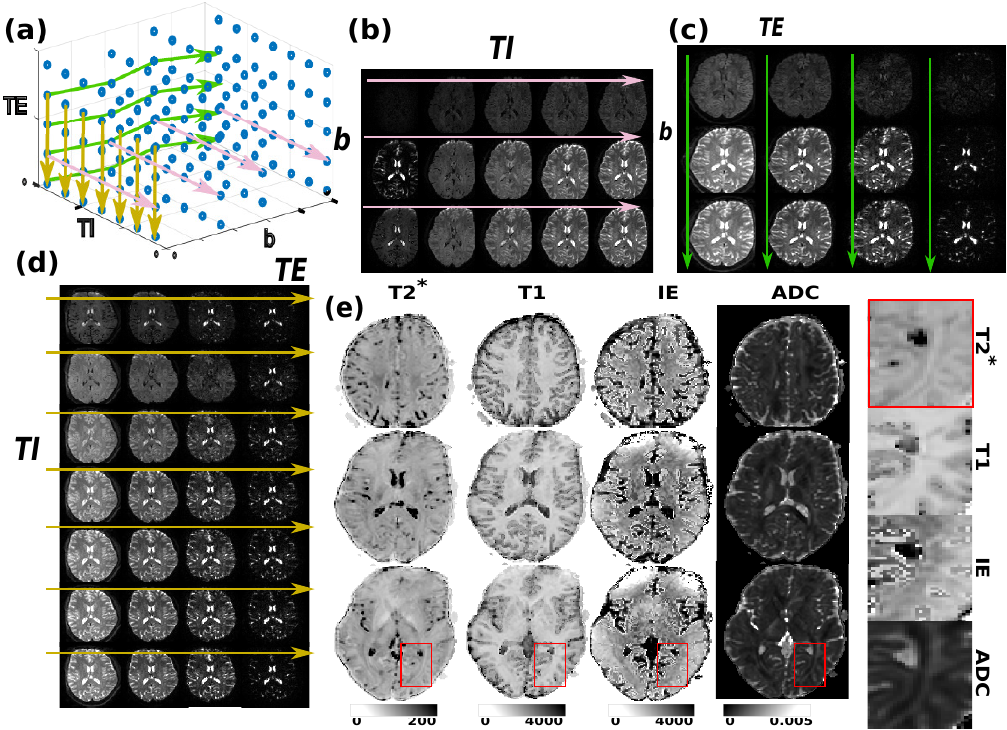}
  \caption{A selected mid-brain slice from an in-vivo three-dimensional data set is depicted. In (a) the obtained sampling in the three-dimensional acquisition parameter space is shown together with three potential cuts to visualize the data content: (b) the TE-TI plane at b=0, (c) The TI-b plane at second echo time and (d) the TE-b plane at longest TI, are depicted. Notice the effect of the interleaved sampling in (b), the lines TI vs. b are not straight lines, but follow the path of the chosen interleaving. (d)The obtained T2$^*$, T1, ADC and inversion efficiency maps are shown at three axial locations in the brain together with a zoom. Note that no post-processing was performed to illustrate the ZEBRA results - particularly no distortion correction.}
  \label{fig:1902_data}
\end{figure}

\begin{figure}[t]
  \centering
  \scriptsize
  \includegraphics[width=0.9\textwidth]{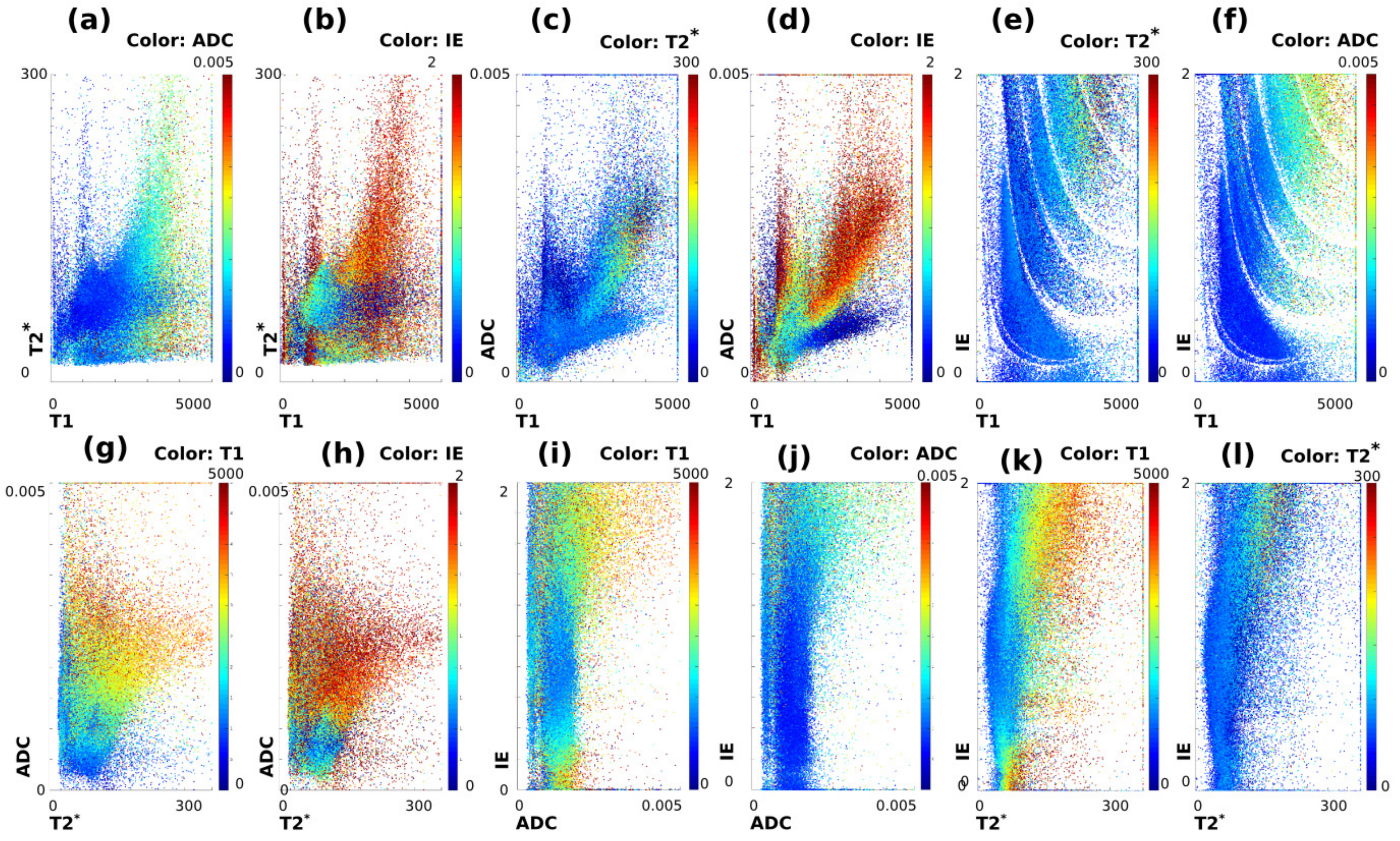}
  \caption{Scatterplots of all voxels within the brain mask illustrating the relations between the four discussed tissue parameters T1, T2$^*$, ADC and Inversion Efficiency (IE). (a)-(b) ADC-T1 space, color decided by (a) inversion efficiency and (b) T2$^*$.  (c)-(d) T1-T2$^*$ space, colored by (c) inversion efficiency and (d) ADC. (e)-(f) T1-Inversion efficiency space, colored by (e) T2$^*$ and (f) ADC. (g)-(h) ADC-T2$^*$ space, colored by (g) T1 and (h) inversion efficiency. }
  \label{fig:scatterplots}
\end{figure}

\begin{figure}[t]
  \centering
  \scriptsize
  \includegraphics[width=0.8\textwidth]{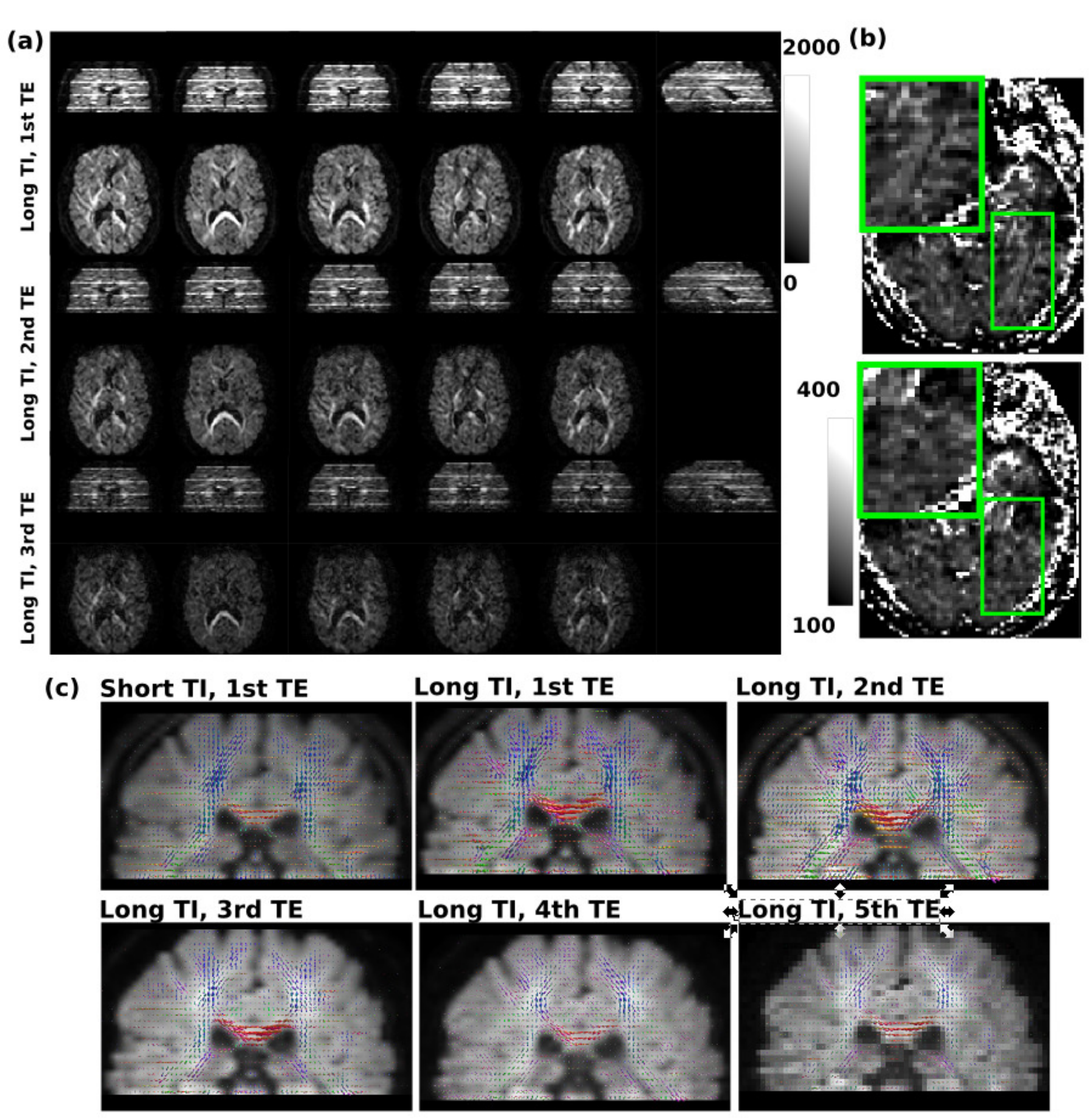}
  \caption{The acquired HARDI-ZEBRA data is shown together with some first preliminary analysis results. (a) The five first volumes are shown as acquired in transverse and in reformatted coronal planes for the first three echos. This represents about 2\% of the total acquired samples. The characteristic ZEBRA look in the reformatted coronal plane is due to $b$-value interleaving. (b) shows T2$^*$ maps calculated for specific directions -- exploiting the availability of the entire parameter space. (c) illustrates the obtained odfs overlayed over the $\ell=0$ spherical harmonics term for both short and long TIs as well all five echo times.}
  \label{fig:odf2}
\end{figure}

\noindent The results from the first adult experiment (E4) are shown
in Fig.  \ref{fig:1902_data} - Fig. \ref{fig:odf2} and Supporting Figure S3.\\ 

\noindent Supporting Fig. S3a shows the coronal view of transverse acquired slice stacks for the first superblock (10 volumes) of the first echo, illustrating the varying TIs and interleaved b-values. Supporting Fig. S3b shows exemplary slices with their corresponding TIs and b-values illustrating in more detail the achieved densely packed data content.\\

\noindent A mid-brain slice from a complete dataset over all TEs, TIs and b-values is depicted in Fig. \ref{fig:1902_data} - (e) shows the obtained T2$^*$, T1, inversion efficiency and ADC maps for subject 1 together with a zoom into the optic radiation. The total acquisition time for this dataset was 2:42min. The estimates of the joint fit are within the expected tissue parameter ranges. No processing or correction for motion and distortion was performed to minimize additional steps from the ZEBRA acquisition to the final paramaters.\\

\noindent The T1, T2$^*$, ADC and Inversion Efficiency values obtained from the adult ZEBRA dataset are displayed in scatter plots in Fig. \ref{fig:scatterplots} for all voxels within the brain mask. The multi-dimensional tissue tissue space is presented for each possible pairing of two acquisition parameters e.g. T1-T2$^*$ in (a,b), T1-ADC in (c,d), T1-IE in (e,f), T2$^*$-ADC in (g,h), IE-ADC in (i,j) and T2$^*$-IE in (k,l). In each case a third tissue parameter is indicated by colour coding the data points. Thus there are two different colored plots for each tissue parameter paring. The scatter plots show the complex and rich inter-relationships between the different tissue parameter.\\

\noindent Results from the second adult protocol (E5) are depicted in Fig. \ref{fig:odf2}. The described HARDI-ZEBRA scan resulted in a total of $42 \cdot 4 \cdot 4=672$ volumes acquired in a total scan time of 22:40 (TR=8s). This corresponds to an acceleration of $N_e N_i = 28$, compared to the next-fastest slice-shuffling approach for combined diffusion-T1 analysis acquired in separate acquisitions with different echo times to sample the TE acquisition parameter space.\\

\noindent The Figure \ref{fig:odf2} illustrates the HARDI-ZEBRA experiment. The first five volumes for the first three echos are depicted - corresponding roughly to 2\% of the acquired samples. Numerous analysis techniques to exploit the data are possible, only two examples are given with (b) directional T2$^*$ maps and (c) TI and TE dependant ODF results, obtained with MRTrix3.

\section*{Discussion}
This paper presented ZEBRA - an integrated acquisition for quantitative MRI of multiple parameters, maximizing efficiency by optimally sharing preparation periods.\\

\noindent The datasets generated are available from the corresponding author on request together with the employed post-processing and fitting scripts.

\subsection*{Flexibility and combination}
\noindent The flexible nature of the proposed ZEBRA approach allows the acquisition parameters (interleaving factor, number of super-blocks, number of acquired echoes) to be tuned to the expected SNR and range of variation in the tissue parameters to be estimated . Using the acquisition parameters as presented allowed for a 20-fold/28-fold decrease in imaging time compared to the separate acquisition of T1-diffusion with SS and T2$^*$-diffusion data.\\

\noindent The proposed T1-diffusion sampling combines effortlessly with acceleration techniques such as SENSE and Half-scan as the proposed conceptual change is independent of the individual EPI read-out. Combination with simultaneous slice imaging techniques is also straightforward providing  further increases in coverage without increasing acquisition time.\\

\noindent The obtained ZEBRA data was also used to obtain the tissue parameter inversion efficiency, and thus was shown to be able to provide a window on magnetization transfer effects. However, further investigation of this parameter is beyond the scope of this paper but can be of interest for future studies \cite{Teixeira2018}.

\subsection*{Additional acceleration by reduction of thermal load}
An additional benefit is that both changes, mixing the diffusion weighting shot-by-shot and adding additional echoes, can decrease the thermal demands on the gradient system, allowing more efficient operation. This is, however, beyond the scope of this study.

\subsection*{Considerations regarding acquisition parameters}
\noindent The T2$^*$-diffusion sampling by multiple echoes leads to additional important considerations for the EPI read-out. Reducing the first echo time to ensure that sufficient signal is maintained to sample the T2* decay curve is in direct competition with high-b value gradient preparations and low sense factors.\\

\noindent Further considerations include the sampling of the TE-TI space: More echoes, as required for higher sensitivity to the longer T2$^*$ range, lead to a less dense TI spacing. This forms an important part of the sampling design, such as, for example, be counteracted by changing the employed interleaving factor.\\

\subsection*{Motion robustness}
\noindent The interleaving of different b-values results in the spreading of the slices forming one volume for same bval/\textbf{bvec} and TI combination across multiple acquired slice stacks. This can lead to more challenging motion patterns within any given volume with a particular weighting. However, as the entire dataset with all acquisition parameter settings forms the input for the multi-parametric modelling, any motion during the acquisition needs to be dealt with independent of interleaving. The more complex motion patterns after sorting the data to individual volumes might contribute to a need for motion correction on the slice level rather then volume level, such as has been used for fetal applications of diffusion  MRI (e.g. \cite{Rousseau2006}) and more recently for adult brain imaging studies \cite{Andersson2016}.  The multi-echo sampling of the required data for T2$^*$ mapping acquires all TE samples within ~200ms and thus effectively freezes motion.

\subsection*{Limitations of this study}
\noindent All experiments and simulations were limited to regularly spaced and interleaved bval/\textbf{bvec} combinations. The sequence however, allows for free distribution of samples to the inversion curve. Further investigation in this direction could be beneficial and will be part of future work.\\
\\
\noindent The focus of this study lies on the presentation of the obtained highly efficient novel acquisition, not on the processing and modelling. The presented joint fits are merely first attempts to explore the obtained data. While the second performed adult scan demonstrated the ability of ZEBRA to deliver a multi-shell HARDI dataset, the directional information and the multi-compartmental origins of the signal were ignored in model~\eqref{eq:signaldecay}. This was a deliberate choice to simplify the multiparametric signal fitting. Future work can focus on advanced multi-dimensional modelling tailored to the rich datasets acquired with the proposed ZEBRA technique.

\subsection*{Outlook}
\noindent Future work will focus on deploying the proposed ZEBRA acquisition for in depth brain studies. In particular, this novel accelerated acquisition ideally lends itself to bespoke multi-compartment models that can disentangle myelin and axonal compartments. Furthermore, ZEBRA has potential for applications outside the brain, such as kidneys, liver or prostate studies.

\section*{Conclusion}
The proposed ZEBRA acquisition combines several novel elements to allow highly efficient integrated sampling of the multi-dimensional acquisition parameter space TE-TI-bval/\textbf{bvec} to obtain quantitative data in the T2$^*$-T1-diffusion domain. Its robustness was shown on phantom and adult brain experiments, the analysis of the several dimensions per voxel shows the independence of the tissue parameters and thus the benefits of this higher dimensional approach. It answers two key challenges, scanning efficiency and consistent, simultaneously acquired data and might thus allow in the future the wider use of joint relaxometry-diffusion techniques both for research applications and clinical use.

\section*{Methods}

\subsection*{Interleaving diffusion encodings}
\begin{figure}[t]
  \centering
  \scriptsize
  \includegraphics[width=0.75\textwidth]{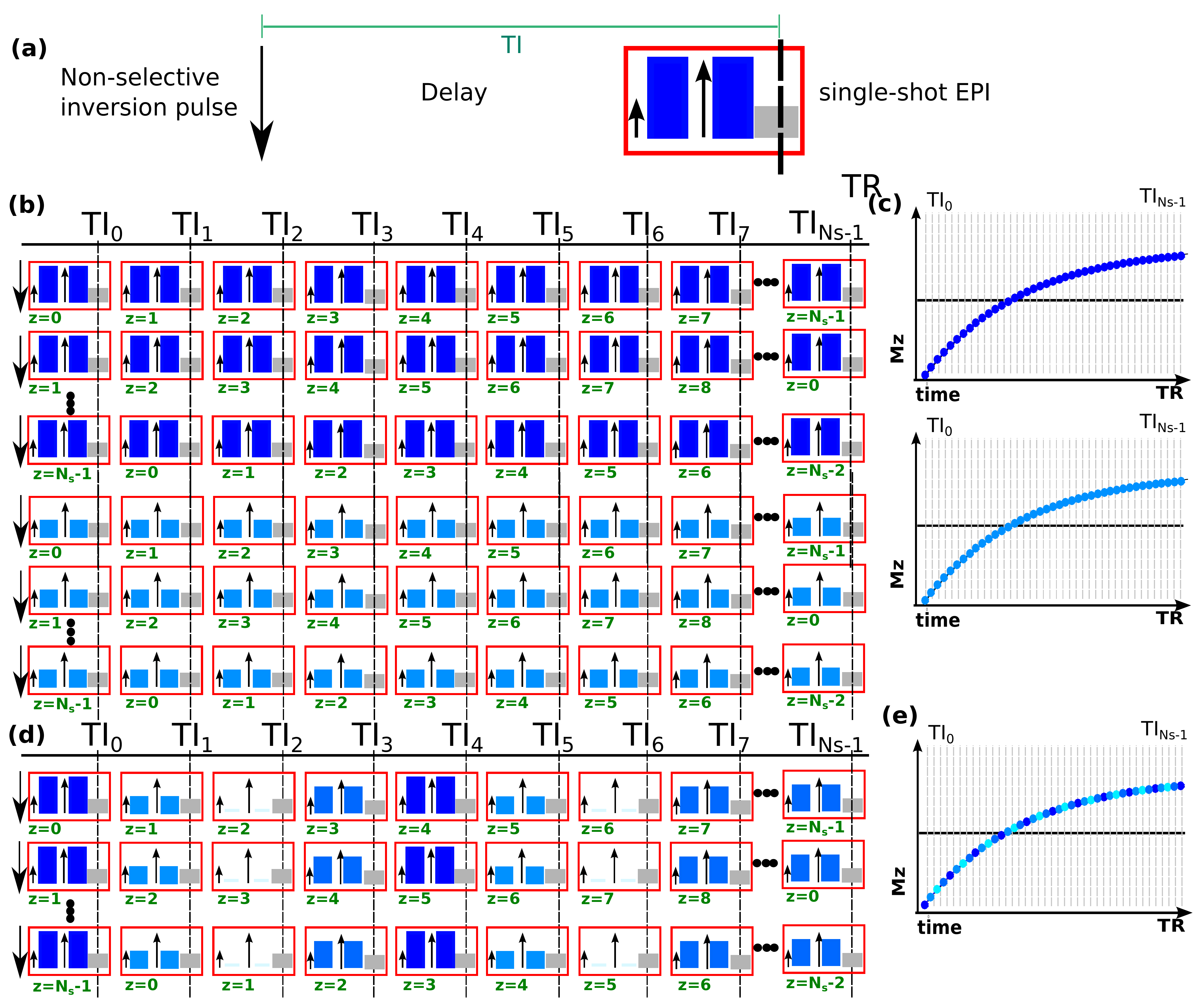}
  \caption{The structure of one single inversion recovery experiment is illustrated in (a), consisting of a global non-selective inversion pulse, a delay time defining the obtained IR contrast and the single-shot EPI module. The IR-dMRI sequence is composed of these blocks as illustrated in (b) and (d). Thereby echo column represents one time sequence within the TR, each row stands for a novel repeat (volume). (c)-top The schematic slice-shuffled acquisition without diffusion interleaving is illustrated for two \textbf{bvec}/bval combinations (highlighted as different shades of blue). (c)-bottom The resulting sampling of the inversion curve using this conventional approach is shown. (d) The schematic slice and diffusion shuffled acquisition schema is illustrated for four \textbf{bvec}/bval combinations. (e) The interleaved sampling of 4 different diffusion encodings along the inversion recovery curve originating from the scheme in (d) is illustrated.}
  \label{fig:seq}
\end{figure}

The conventional IR experiment consists of a slice selective inversion pulse, a delay time defining the TI and a single- or multi-shot read-out as illustrated in Fig. \ref{fig:seq}a for the case of single shot EPI (ssEPI). SS techniques use a non-selective inversion pulse followed by a stack of ssEPI slices excited in such a way that any preparation time required to achieve a specific TI in one slice is used to sample another slice at a shorter TI. This leads to one continuous acquisition and no waiting delay times as illustrated in Fig. \ref{fig:seq}b. If there are $N_s$ slices, then the whole procedure must be repeated $N_s$ times with the slice firing order sequentially shifted to achieve a set of complete stacks . Each slice is also diffusion weighted and they require every slice ($s=1,..,N_s$) to be sampled with every inversion time (TI) for every diffusion encoding ($d=1,..,N_d$), where each $d$ signifies one combination of bval/\textbf{bvec}. The total acquisition thus consists of $N_s\times N_d$ volumes, parametrized by volume number $v$ with $v=1,..,N_s\times N_d$. This approach is extremely inefficient as the inversion recovery curve is generally massively oversampled: typical values of TR=6s and $N_s=40$ lead to data points at 150msec intervals (Fig. \ref{fig:seq}c) with the minimum time determined by the EPI shot time. More appropriate, sparser sampling is not possible due to the limitations of the traditional 'one volume – one diffusion encoding' scheme where every slice is sampled with the same diffusion encoding before the next encoding is chosen.\\

\begin{figure}[t]
  \centering
  \scriptsize
  \includegraphics[width=0.8\textwidth]{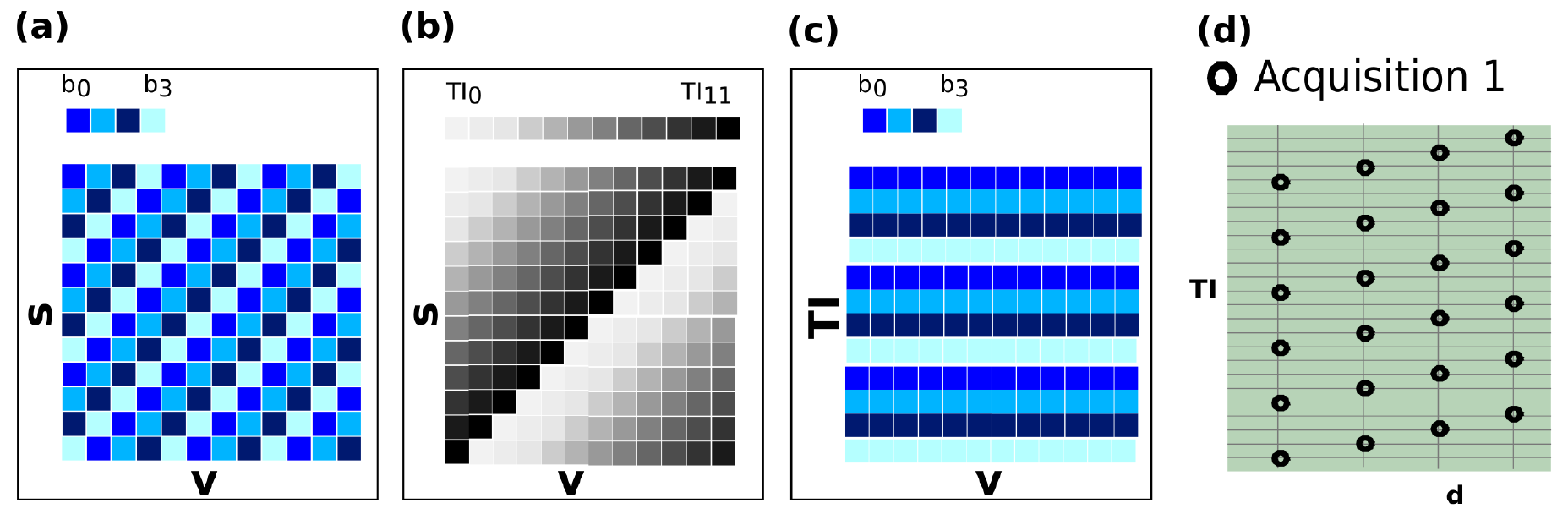}
  \caption{The chosen mappings showing (a) the employed diffusion encodings and (b) the inversion times mapping from volumes v to geometric location s. In (c), the resulting mapping of diffusion encoding in the volume v vs. inversion time TI plane is shown. (d) Shows the acquired sampling in the TI vs. d plane.}
  \label{fig:orders}
\end{figure}

\noindent Our approach breaks with this traditional paradigm by changing the diffusion encoding per slice \cite{Hutter2017}. By interleaving $N_i$ different diffusion encodings within the TR (Fig. \ref{fig:seq}d, where $N_i=4$), multiple diffusion encoding combinations can be sampled within a single recovery curve (Fig. \ref{fig:seq}e). Instead of repeating the same diffusion encoding with all $N_s$ slice shuffles, this technique only requires $N_s/N_i$ repeats. To combine the needs to sample every slice with every diffusion encoding and to achieve full diffusion encoding volumes per TI, a rolling approach is chosen: The diffusion encoding alternates between the chosen $N_i$ encodings within every volumes and  the start of this permutation is incremented by one for subsequent volumes. Fig. \ref{fig:orders}a and b show the chosen diffusion encoding (a) and inversion times (b) in the $s$ vs. $v$ space. The four colors in Fig. \ref{fig:orders}(a) represent four separate bval/bvec combinations illustrating the shifted permutations from volume to volume. The grey levels in Fig. \ref{fig:orders}(b) represent all chosen TIs and their incrementation from volume to volume. This leads, as illustrated in Fig. \ref{fig:orders}c ($TI$ vs. $v$) to a constant mapping of $d$ to $TI$. The sampling in the T1-diffusion domain, shown in Fig. \ref{fig:orders}d is thus on an oblique lattice with flexible control of sampling density.\\

\noindent The order of the chosen $N_i$ encodings is not relevant for the ability to obtain accelerated IR-dMRI scans, it was, however, chosen such that the low-b volumes are maximally spread over time to enhance the potential for further motion correction \cite{Hutter2017}.

\subsection*{Super-block design}
We use the term super-block to refer to each sequence of $N_s$ volumes. While all super-blocks are equal in length, they differ in the number of included interleaved diffusion weightings $N_i$ for this block. While the same TIs on the recovery curve are sampled in every super-block, varying the length $N_i$ - and thus the number of combined diffusion encodings – allows different sampling densities for each super-block. For these IR-dMRI blocks, the length $N_i$, and the slice number $N_s$ define the number of points on the recovery curve per diffusion encoding as $N_{TI}=N_s/N_i$. Superblock $N_i$ can for example be selected to balance the declining SNR at higher diffusion weightings, recognizing that lower b-values can allow robust T1 estimation with fewer points on the recovery curve.\\
\\
\noindent Finally, super-blocks without a pre-inversion - termed dMRI super-block - can be included into the sequence to achieve additional dMRI samples at a maximum data rate – these would, however, not be suitable for the simple combined modelling approaches. The sampling points for those equals $N_{TI}=1$ and the length of these dMRI blocks equals $N_v=N_i$.\\
\noindent The final total acquisition will be made up by $N_b$ (variable length) super-blocks of total length $\sum_b^{N_b} N_v^b$.\\

The choice of $N_i=4$ is shown in Fig. \ref{fig:seq}d and e, additional choices with more details are depicted in Supporting Figure S1 for $N_i=8,4,2$ next to a dMRI super-block (rightmost column) and the composition of the total acquisition (d).

\noindent The sequence structure and TR remains the same for both IR-dMRI and dMRI blocks, but an inversion pre-pulse is only played out for IR-dMRI super-blocks.

\subsection*{Integrated T1-T2*-Diffusion sampling}
\begin{figure}[t]
  \centering
  \scriptsize
  \includegraphics[width=0.9\textwidth]{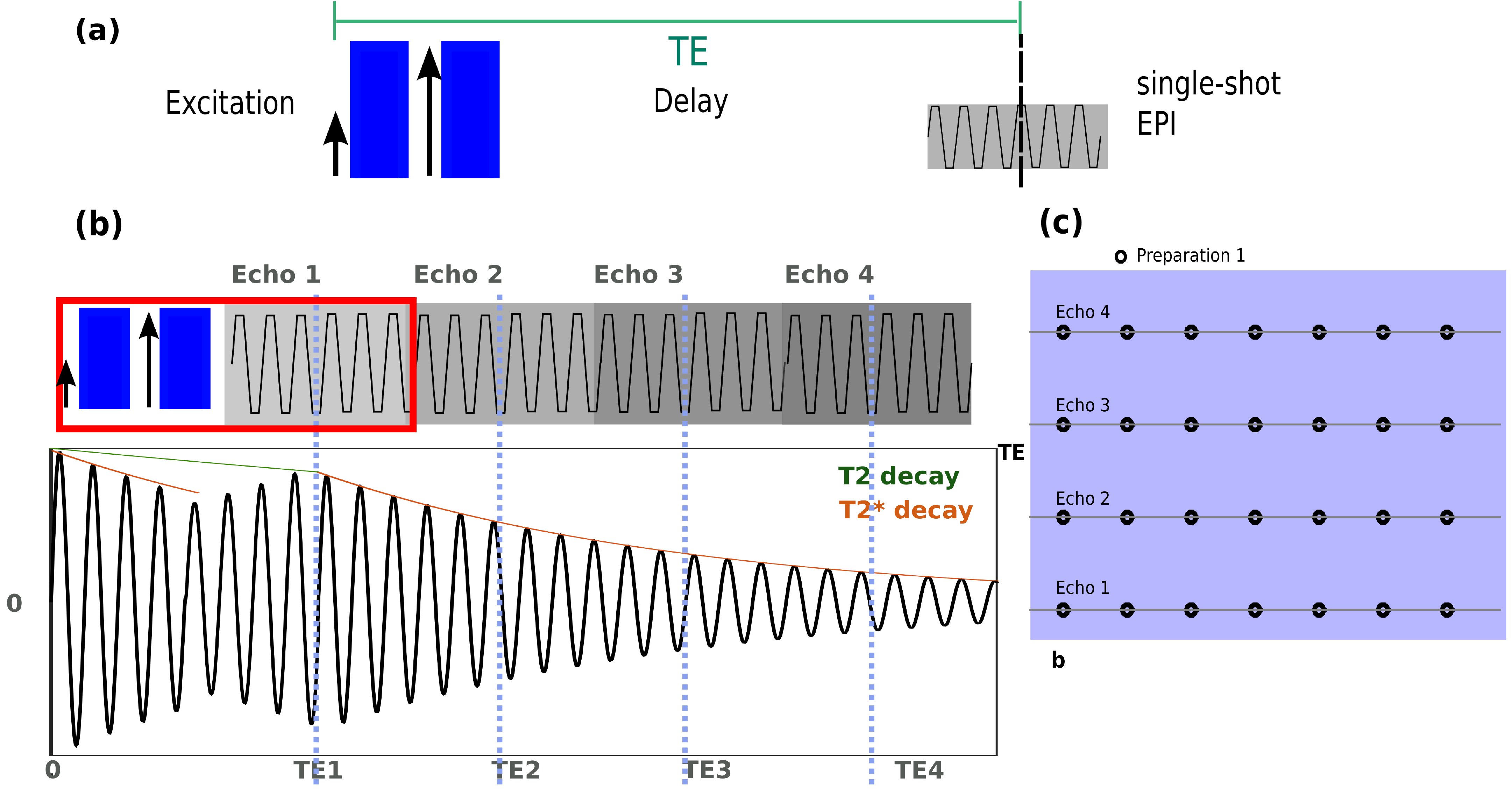}
  \caption{The employed multi-echo Spin-echo Gradient Echo scheme is depicted schematically in (a) together with the resulting sampling in the TE-b plane in (b). The red box in (b) depicts a conventional dMRI experiment. Finally, (c) describes the obtained sampling in the b-TE plane with the proposed acquisition.}
  \label{fig:muechos}
\end{figure}

\noindent The required echo times for T2/T2$^*$ estimation are typically obtained by including a delay between excitation and read-out to vary the echo time as illustrated in Fig. \ref{fig:muechos}a.\\ Especially for the sampling of the higher TEs, these idle times render the experiment inefficient. ZEBRA employs a multi-echo approach to efficiently use all preparation times. Instead of repeating the diffusion preparation in different acquisitions by varying the TE, one acquisition employs one diffusion preparation to sample multiple TEs. Therefore, SE-EPI sequence was extended to a multi-echo sequence by adding several EPI read-out blocks with minimal spacing after the initial diffusion preparation and SE-read-out block as illustrated in Fig. \ref{fig:muechos}. The $d$-TE space is thus sampled simultaneously with the subsequent echoes. The extension from the spin-echo sequence to a double echo SE-GE sequence was previously exploited for dynamic distortion correction \cite{Cordero-Grande2018}.

\subsection*{Post-processing}
The data is sorted to individual diffusion-encoding volumes at the obtained $N_s/N_i$ TIs and all TEs. Each resulting volume is thus parametrized by [\textbf{bvec} bval TE and TI] and the set of all volumes forms the input for any further modelling step. In the following, the data in every voxel was fitted to a joint signal model to obtain the most basic tissue parameters proton density $S_0$, $T1$, $T2^*$ and $ADC$ using the following equation
\begin{equation}
\label{eq:signaldecay}
S=S_0 (1-IE\exp^{-(TI/TE)/T1}+\exp^{-(TR/T1)})\exp^{-b ADC}\exp^{-(TE/T2^{*})}.
\end{equation}
An inversion efficiency ($IE$) is is also estimated in order to account for incomplete inversion (full inversion corresponds to $IE=2$). It is known that a bi-exponential longitudinal relaxation is present in tissues with strong magnetization transfer (MT) effects, such as white matter \cite{vanGelderen2016}. MT effects are expected to alter the estimated values of both $S_0$ and $IE$. 

\noindent Tissue parameter estimates of $S_0$, $IE$, $T1$, $ADC$ and $T2^*$ were obtained based on a least-square criteria between Eq. \ref{eq:signaldecay} and data sampled via the proposed ZEBRA acquisition. For each voxel individually, tissue parameter estimates were found using MatLab 2014b Levenberg-Marquart algorithm, with a function tolerance value of $10^{-9}$ and a fixed initial guess set to $S_0=4000ms, IE=2, T1=1000ms, T2*=200ms, ADC=0.003mm^2s^{-1}$.

\section*{Acknowledgements}
This work received funding from the ERC (FP7/20072013)/ERC no. 319456 (dHCP), the Human Placenta Project (NIH 1U01HD087202-01), the Wellcome Trust (201374/Z/16/Z) and was supported by the Wellcome EPSRC CME(WT 203148/Z/16/Z), MRC (MR/K006355/1) and by the NIHR Biomedical Research Centre. The views expressed are those of the authors and not necessarily of the NHS, NIHR or the Department of Health.

\section*{Author contributions statement}
Ja.Hu. and Jo.Ha. conceived the experiments. J.H. and A.P. implemented the method, R.T. and S.M. informed the analysis. J.H. and P.S. and L.J. conducted the experiments, J.H., D.C., T.R. and S.M. analysed the data. All authors reviewed the manuscript.

\section*{Additional information}
The author(s) declare no competing interests.

\bibliography{invT1}

\end{document}